\documentclass[fleqn,12pt,twoside]{article}
\usepackage[headings]{espcrc1}

\readRCS
$Id: espcrc1.tex,v 1.2 2004/02/24 11:22:11 spepping Exp $
\usepackage{graphicx}
\usepackage[figuresright]{rotating}

\newcommand{\AmS}{{\protect\the\textfont2
  A\kern-.1667em\lower.5ex\hbox{M}\kern-.125emS}}

\title{Measurements of the nuclear modification factor and
elliptic flow of $\phi$ mesons at RHIC}

\author{X. Cai\address[SINAP]{Nuclear Physics Department, Shanghai Institute of Applied Physics,\\
        P.O. Box 800-204, Shanghai, 201800, China}
        (for the STAR\thanks{For the full list of STAR authors and acknowledgments,
        see appendix 'Collaborations' of this volume.} Collaboration)}

\runtitle{$R_{CP}$ and $v_2$ of $\phi$ mesons at RHIC}
\runauthor{X. Cai}

\begin{document}

\maketitle

\begin{abstract}
The observation of meson and baryon grouping in the $R_{CP}$ and
$v_2$ measurements at intermediate $p_T$ has been interpreted as a
manifestation of bulk partonic matter hadronization through
multi-parton dynamics such as recombination of partons. $\phi$
mesons provide unique sensitivity to test these theoretical
scenarios, since the $\phi$ has a mass heavier than the proton and
close to the hyperons. The $R_{CP}$ and $v_2$ measurements of
$\phi$ mesons from Run IV Au+Au collisions at $\sqrt{s_{NN}}=200$
GeV at STAR are presented. Energy and colliding system dependence
of the $\phi$ yields at mid-rapidity are discussed. The results
are compared to the measurements of other hadrons. Properties of
strange quarks in the bulk matter at hadron formation are
discussed.
\end{abstract}

\vspace{-0.2cm}
\section{INTRODUCTION}
\vspace{-0.2cm}

Strange particle production may be sensitive to the existence and
properties of a deconfined partonic state formed in relativistic
heavy-ion collisions. Current measurements of identified hadrons
by STAR ($K_{s}^{0}$ and $\Lambda$) \cite{STAR1} and PHENIX
(proton and $\pi^{0}$) \cite{PHENIX} show that the nuclear
modification factor ($R_{CP}$) for the $\Lambda$ differs from that
of the $K_{s}^{0}$ and $R_{CP}$ for the proton differs from that
of the $\pi^{0}$, which indicates a dependence on particle species
of particle production. This phenomenon can be explained by quark
coalescence or recombination models \cite{REC1,REC2,REC3}, in
which the hadrons at intermediate $p_{T}$ are predominantly formed
by the coalescence of constituent quarks from a thermalized
partonic system. The $\phi$ meson is of particular interest in
distinguishing between dependence on mass or particle species,
since the $\phi$ meson has a mass similar to that of the $\Lambda$
baryon. Since the $\phi$ interaction cross-section with other
hadrons is small \cite{SHOR}, $\phi$ will retain information from
the early hot and dense phase. Additionally, since kaon
coalescence has been ruled out as the dominant $\phi$ production
mechanism at RHIC \cite{JGMA}, measurements of elliptic flow
($v_{2}$) of $\phi$ should be a sensitive probe for the build-up
of pressure in the early reaction stage of relativistic heavy ion
collisions.

\vspace{-0.2cm}
\section{TRANSVERSE MASS DISTRIBUTION}
\vspace{-0.2cm}

Reconstruction of $\phi$ mesons is accomplished by calculating the
invariant mass ($m_{inv}$) of all possible pairs of $K^{+}$ and
$K^{-}$ candidates in each event for each transverse momentum
($p_{T}$) bin and centrality bin. The combinatorial background is
calculated by using the mixed-event technique \cite{JGMA,EM1,EM2}.
The transverse mass distributions of $\phi$ mesons from Au+Au
collisions (at 62.4GeV, 130GeV \cite{STAR4} and 200 GeV, Run II
and Run IV datasets), d+Au \cite{CAI} and p+p \cite{JGMA}
collisions (at 200 GeV) are measured at STAR. The collisions are
divided into different centrality classes, where each centrality
bin corresponds to a certain fraction of the total hadronic
cross-section. The measured STAR $\phi$ meson spectra for
$\sqrt{s_{NN}}=200$ GeV Au+Au collisions are consistent across the
Run II and Run IV datasets.

\begin{figure}[htb]
\vspace{-0.9cm}
\begin{minipage}[t]{80mm}
\resizebox{!}{6cm}{
\includegraphics[width=5cm]{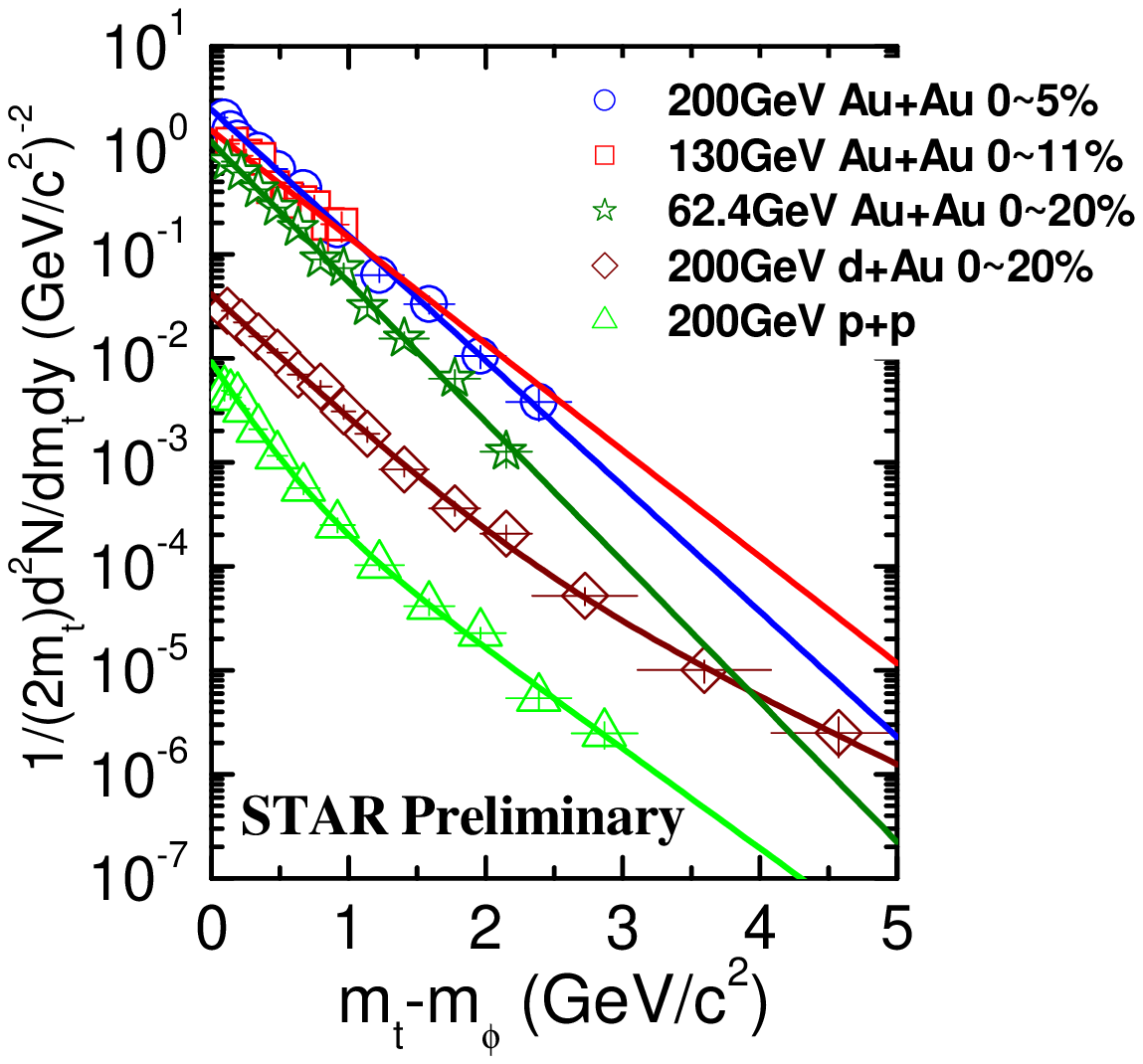}}
\vspace{-0.8cm} \caption{The transverse mass distributions of
$\phi$ mesons for the most central collisions for different
collision systems. Error bars are statistical errors only.}
\label{fig:DouExp}
\end{minipage}
\hspace{\fill}
\begin{minipage}[t]{75mm}
\resizebox{!}{5cm}{
\includegraphics[width=6cm]{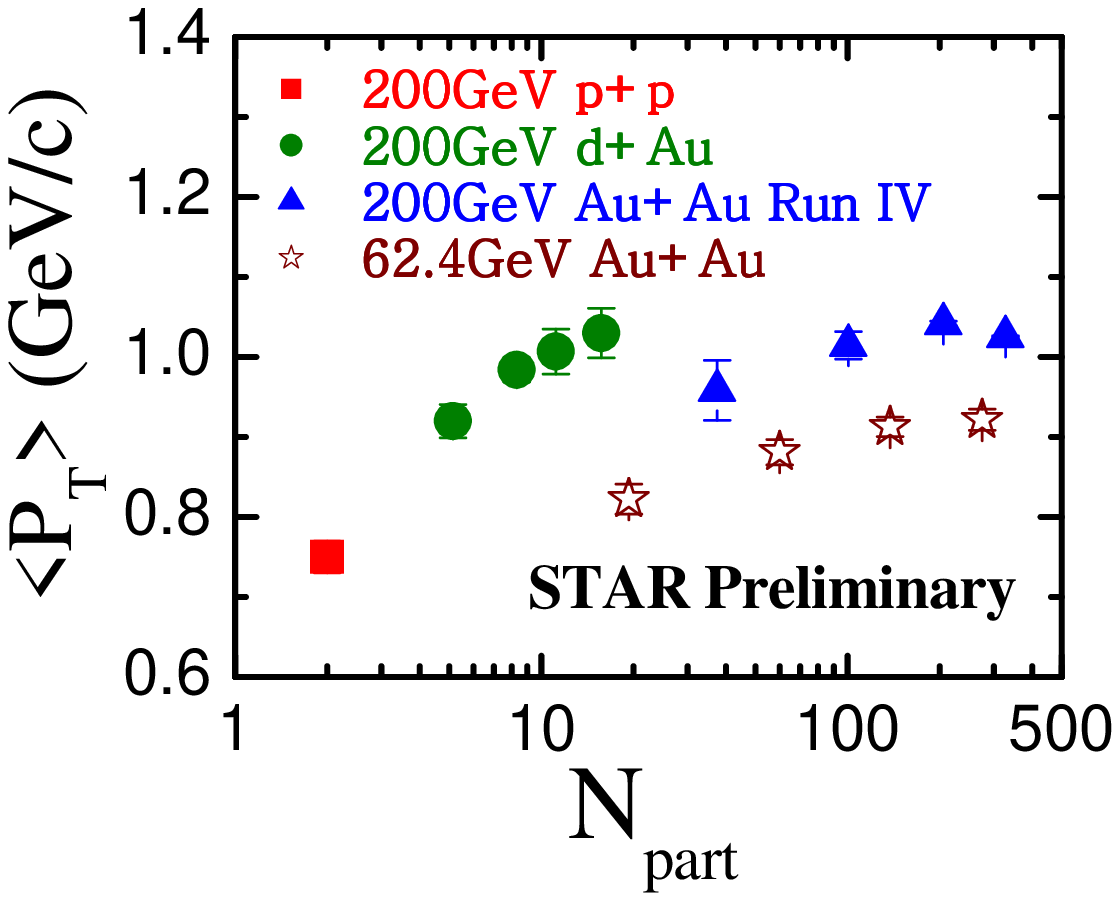}}
\vspace{-0.8cm} \caption{$\phi$ $<p_{T}>$ as a function of
participant number. Error bars are statistical errors only.}
\label{fig:MeanPt}
\end{minipage}
\vspace{-0.9cm}
\end{figure}

In order to compare the spectral shapes, the most central $\phi$
spectra for different collision systems are plotted together in
Figure~\ref{fig:DouExp}. It can be seen that the $\phi$ meson
spectra for Au+Au collisions in the measured $p_{T}$ range can be
described by an exponential function for all collision energies.
The $\phi$ spectra for d+Au and p+p collisions deviate from the
exponential distribution and have a power-law tail in the
intermediate $p_{T}$ range, where the double-exponential function
\cite{CAI} can reproduce the experimental data well. Due to the
limited $p_{T}$ range for Au+Au collisions, measurements at
$(m_{t}-m_{\phi})>$2.5 GeV/$c^{2}$ may be necessary to see if a
power-law tail exists.

The system-size and beam-energy dependence of $\phi$ $<p_{T}>$ is
shown in Figure~\ref{fig:MeanPt}. For different collision systems
at 200 GeV, the $<p_{T}>$ increases from p+p to d+Au collisions as
a function of participant number ($N_{part}$). $\phi$ $<p_{T}>$ in
Au+Au collisions doesn't change significantly within error bars,
unlike the general increasing trend for $\bar{p}, K^{-}$ and
$\pi^{-}$ \cite{STAR5,JGMA}. This is consistent with an early
freeze-out scenario for the $\phi$-meson. If the $\phi$ hadronic
scattering cross-section is much smaller than that of other
particles, one would not expect the $<p_{T}>$ distribution to be
appreciably affected by any final state hadronic rescatterings. It
is also seen that the $<p_{T}>$ of the $\phi$ meson in Au+Au
collisions at 62.4GeV is lower than that at 200 GeV. Since the
$<p_{T}>$ carries information about radial flow, it may be
different at different collision energies.

\vspace{-0.2cm}
\section{NUCLEAR MODIFICATION FACTOR ($R_{CP}$)}
\vspace{-0.2cm}

\begin{figure}[ht]
\vspace{-1cm}
\begin{minipage}[t]{77mm}
\resizebox{!}{6cm}{
\includegraphics[width=5cm,height=4.3cm]{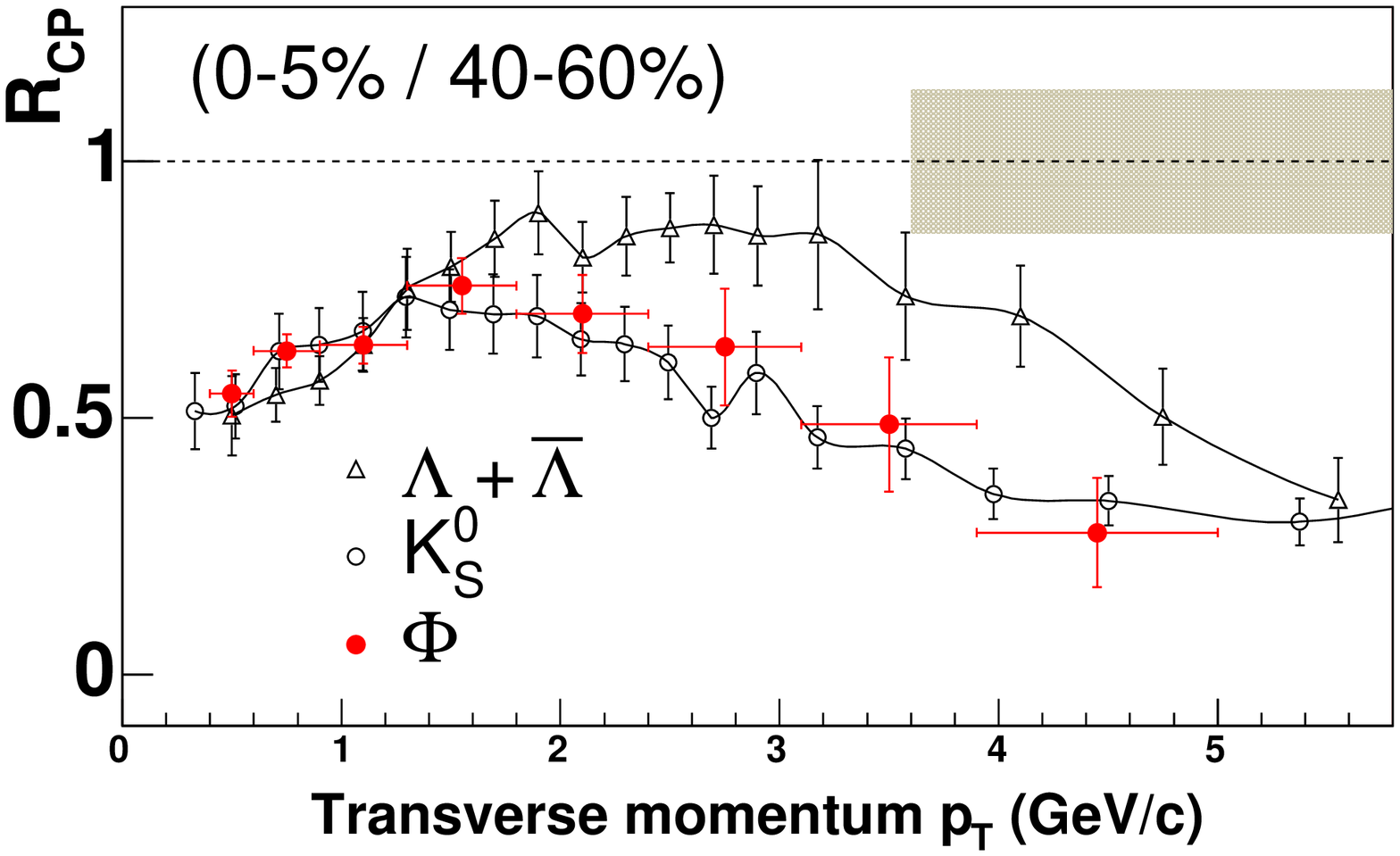}}
\vspace{-0.7cm} \caption{The ratio of yields in central collisions over
peripheral collisions ($R_{CP}$) normalized by $\langle
N_{bin}\rangle$ in Au+Au collisions at 200 GeV.
\label{fig:AuAuRCP}}
\end{minipage}
\hspace{\fill}
\begin{minipage}[t]{77mm}
\resizebox{!}{5.3cm}{
\includegraphics[width=5cm]{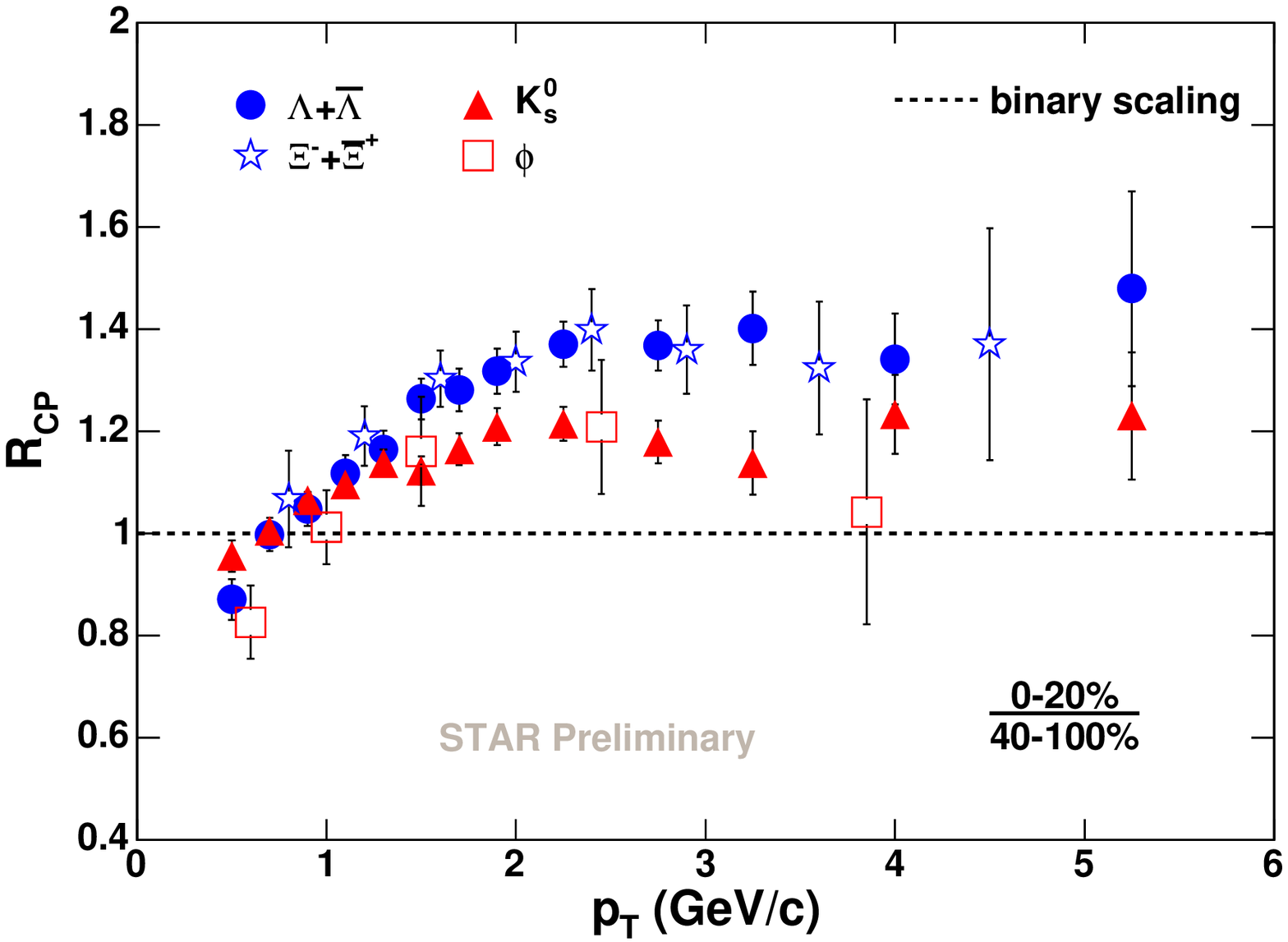}}
\vspace{-0.7cm} \caption{The ratio of yields in central collisions over
peripheral collisions ($R_{CP}$) normalized by $\langle
N_{bin}\rangle$ in d+Au collisions at 200 GeV.\label{fig:dAuRCP}}
\end{minipage}
\vspace{-0.7cm}
\end{figure}

$R_{CP}$ is calculated as the ratio of the yields from central
collisions to peripheral collisions scaled by the mean number of
binary collisions. Comparisons of the $R_{CP}$ for Au+Au
collisions ($K_{s}^{0}, \phi$ and $\Lambda$) and d+Au collisions
($K_{s}^{0}, \phi, \Lambda$ and $\Xi$) at 200 GeV are shown in
Figure~\ref{fig:AuAuRCP} and Figure~\ref{fig:dAuRCP},
respectively. Only statistical errors are included in the figures.

In both collision systems at intermediate $p_{T}$, the $R_{CP}$ of
baryons ($\Lambda$ and $\Xi$) is larger than that of mesons
($K_{s}^{0}$ and $\phi$), which implies that particle production
in this $p_{T}$ region is driven by the particles' types, not
their masses. The $R_{CP}$ results are consistent with the
partonic recombination model predictions \cite{REC1,REC2,REC3}
that the centrality dependence of the yield at intermediate
$p_{T}$ depends more strongly on the number of constituent quarks
than on the particle mass. There also may be a tendency for values
of $R_{CP}$ for all particles to approach each other at high
$p_{T}$. For 200 GeV d+Au collisions, the measurements are
consistent with the proposal by Hwa and Yang \cite{HWA} that the
particle type dependence of the Cronin effect may not be due to
the initial parton scatterings alone. More data are needed for a
firm conclusion.

\vspace{-0.2cm}
\section{ELLIPTIC FLOW OF $\phi$ MESONS IN 200 GeV Au+Au COLLISIONS}
\vspace{-0.2cm}

\begin{figure}[ht]
\vspace{-1cm}
\begin{center}
\includegraphics[width=7cm,height=5cm]{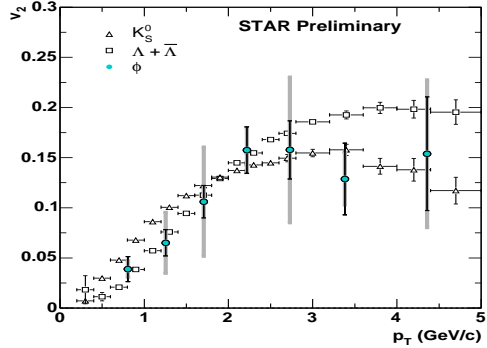}
\vspace{-0.8cm} \caption{The $v_{2}$ parameter for $\phi$ meson as
a funtion of $p_T$. In comparison, the $v_{2}$ for $K_{s}^{0}$ and
$\Lambda$ has also been plotted. Systematic uncertainties from
comparison of two methods are shown as gray bands.\label{fig:V2}}
\end{center}
\vspace{-0.8cm}
\end{figure}

The first measurements of $v_{2}$ for $\phi$ at mid-rapidity in
Run IV 200 GeV Au+Au collisions are presented in
Figure~\ref{fig:V2}. The $v_{2}$ is calculated as $\langle
cos[2(\phi_{i}-\Psi_{RP}^{i})]\rangle$ and the auto-correlation is
subtracted \cite{STAR3}. It can be seen that the $v_{2}$ of $\phi$
increases monotonically for $p_{T}<2.0GeV/c$ and becomes flat in
the intermediate $p_{T}$ range. Within statistical uncertainties,
the minimum-bias $\phi$ meson results are similar to the $v_{2}$
of the $K_{s}^{0}$, which shows number of constituent quarks
scaling \cite{STAR3,CHEN}. However, the error bars in the
intermediate $p_{T}$ range are still large.

\vspace{-0.2cm}
\section{CONCLUSION}
\vspace{-0.2cm}

In summary, the measurements of $\phi$ transverse mass
distributions at mid-rapidity from different collision systems at
RHIC are reported. It is found that the shape of the $\phi$
spectra for Au+Au collisions in a limited $p_{T}$ range can be
described by an exponential function, while the $\phi$ spectra of
d+Au and p+p collisions have a power-law tail in the intermediate
$p_{T}$ range and the double exponential function fits the data
well. The $R_{CP}$ for $\phi$ mesons in 200 GeV d+Au collisions
shows the Cronin effect as seen in low energy p+A collisions
\cite{CRONIN}. The $R_{CP}$ measurements are divided into two
groups in both 200 GeV d+Au and 200 GeV Au+Au collisions in the
intermediate $p_{T}$ range, where the $R_{CP}$ of baryons
($\Lambda$ and $\Xi$) is larger than that of mesons ($K_{s}^{0}$
and $\phi$). This particle species dependence of $R_{CP}$ will
constitute a unique means to investigate the hadronization
mechanism of the dense matter formed in nucleus-nucleus
collisions. The measurements of $v_2$ for $\phi$ mesons in Run IV
200 GeV Au+Au collisions at STAR are also presented. It is seen
that the $v_2$ of $\phi$ meson is similar to that of $K_{s}^{0}$.
Since the $\phi$ meson is not produced via $K^+K^-$ fusion
\cite{JGMA}, this implies partonic collectivity at RHIC.

\vspace{1.0cm}

{\it Acknowledgments---} Besides the acknowledgments in appendix,
we wish to thank NSFC 10475108 and 03 QA 14066 for contributing to
the travel expenses.

\end{document}